\documentclass[runningheads,a4paper]{llncs}

\usepackage{amssymb}
\usepackage{graphicx}

\usepackage{url}
\urldef{\mailsa}\path|{alfred.hofmann, ursula.barth, ingrid.haas, frank.holzwarth,|
\urldef{\mailsb}\path|anna.kramer, leonie.kunz, christine.reiss, nicole.sator,|
\urldef{\mailsc}\path|erika.siebert-cole, peter.strasser, lncs}@springer.com|    
\newcommand{\keywords}[1]{\par\addvspace\baselineskip
\noindent\keywordname\enspace\ignorespaces#1}

\usepackage{listings}
\usepackage{wrapfig}
\usepackage{gensymb}
\usepackage[utf8]{inputenc}
\usepackage{paralist}
\usepackage{color}
\usepackage{ifthen}

\newcommand{\clcaption}[2]{
  \ifthenelse{\equal{#2}{}}{\caption[#1]{\textbf{#1}}}{\caption[#1]{\textbf{#1}: #2}}
}

\newcommand{\figuremacro}[4]{
  \begin{figure}[htbp]
    \medskip 
    \centering
    \includegraphics[width=#4\columnwidth]{#1}
    \clcaption{#2}{#3}
    \label{#1}
  \end{figure}
}

\definecolor{dkgreen}{rgb}{0,0.6,0}
\definecolor{gray}{rgb}{0.5,0.5,0.5}
\definecolor{mauve}{rgb}{0.58,0,0.82}

\graphicspath{{img/}}

\definecolor{Red}{rgb}{1,0,0}

\definecolor{Orange}{rgb}{1,0.5,0}

\begin{document}

\mainmatter  

\title{FreeIMU: An Open Hardware Framework for Orientation and Motion Sensing}

\titlerunning{ }

%
%
\author{Fabio Varesano}
\authorrunning{ }

\institute{Dipartimento di Informatica, Univ. Torino\\
C.so Svizzera 185, 10149 Torino, Italy\\
\url{http://www.di.unito.it/~varesano}}

%
%

\toctitle{Lecture Notes in Computer Science}
\tocauthor{Authors' Instructions}
\maketitle

\begin{abstract}
Orientation and Motion Sensing are widely implemented on various consumer products, such as mobile phones, tablets and cameras as they enable immediate interaction with virtual information. The prototyping phase of any orientation and motion sensing capable device is however a quite difficult process as it may involve complex hardware designing, math algorithms and programming.

In this paper, we present FreeIMU, an Open Hardware Framework for prototyping orientation and motion sensing capable devices. The framework consists in a small circuit board containing various sensors and a software library, built on top of the Arduino platform. 
Both the hardware and library are released under open licences and supported by an active community allowing to be implemented into research and commercial projects.
\end{abstract}

\section{Introduction}

In the last five years, with the introduction of mainstream products such as the Apple iPhone or the Nintendo Wii, supported by remarkable advances in the sensors technology, orientation- and motion sensing-based interaction has become very common among consumer products. Such interaction is currently implemented in many products such as mobile phones, digital cameras, camcorders, tablets, etc. and it has been also used as key technology in various research projects ranging from HCI to Medical devices.


The complexity behind implementing even simple orientation and motion sensing capable devices is however quite remarkable as it may require developing ad-hoc hardware  supporting the sensors needed, understanding and implementing complex sensor fusion algorithms as well as developing embedded software for sensor reading. Such complexity makes the process of prototyping orientation and motion sensing capable devices time consuming, difficult, costly as well as difficult to maintain. 

This is probably the reason why much of the existing literature in HCI using some kind of orientation sensing relies on commercially available devices. The Nintendo Wiimote, Apple iPhone and iPod Touch as well as Android phones or Playstation Move controllers, have been widely used as prototyping devices in orientation sensing and accelerometer based applications. Some of such works include \cite{Lee:2008:WCC:1347390.1347400}, \cite{Liarokapis:2009:MAR:1667552.1667557}, \cite{Folmer:2012:SGU:2148131.2148161} as well as MIT Funf Sensing Framework\footnote{\url{http://funf.media.mit.edu/}}. The Wiimote has even been the base of the \textit{Wiimote hackery workshop} held at TEI 2010 \cite{Williams:2010:WHS:1709886.1709976}.

In our opinion, even if such devices proved to be effective in some prototyping applications, there are many limits in specific implementations. The biggest limitation is the fact that they can't be easily used to prototype other devices or controllers. Using something like the Wiimote or iPhone poses various limitations in the size, features and costs of the prototype. This is a remarkable limitation when working on specific HCI devices in which the shape is to be evaluated during the prototyping. An example might be ad-hoc controllers used as physical representation of virtual information as part of Tangible User Interfaces. In such applications, the controllers have various constraints in shape and size since they need to represent virtual information. Commercial devices generally can't be extended with other input/feedback technologies: even adding simple feedback capabilities, like for example an RGB light, can be challenging when using closed commercial products.

This has been partially solved by the introduction of commercial sensor boards, usually designed for stand-alone operation or to be interfaced with microcontroller-powered prototyping boards, such as Arduino. However, even if these boards solve some of the hardware complexity of prototyping orientation and motion sensing devices, they are still limited by the complexity of sensor fusion algorithms and programming. Some vendors provide their own sensor fusion firmware, however they usually release it as closed source software, thus extremely complex to embed in ad-hoc designed applications or to be extended.

In this paper we propose the FreeIMU framework developed to solve the need of rapid prototyping of orientation and motion sensing capable devices. In the paper we briefly describe the FreeIMU framework, its hardware and libraries and then provide examples of it usage. References to related works are provided as well as a discussion of future developments.

\section{The FreeIMU framework}

The FreeIMU framework is composed by a small sensor board and a software library. The project has been developed as an Open Hardware with the contribution of a community of developers.\\The FreeIMU hardware consists in a very small board containing 3-axis accelerometer, gyroscope and magnetometer in a 9 Degrees of Measurement (DOM) Magnetic Angular Rate Gravity (MARG) Inertial Measurement Unit (IMU). Later versions also feature an high resolution barometer.\\The board can be used with the FreeIMU library, a set of easy to use APIs implementing sensor fusion algorithms on top of the Arduino platform\cite{arduino_chi}, allowing the developer to easily obtain a precise 4 Degrees of Freedom (DOF) sensing device which is then capable of sensing its own yaw, pitch, roll and altitude with respect to the external world. When the tracking precision doesn't need to be accurate, it is also possible to track the complete motion of the device in 6 DOF.

Starting from the FreeIMU framework is easy to prototype orientation and motion sensing capable devices which use a FreeIMU compatible board connected to an Arduino. Such setup, thanks to the flexibility of the Arduino platform, also allows to be easily extended with more sensors (eg. ambient light or proximity sensors), other input/output hardware (eg. buttons, joysticks, vibration motors) and communication technologies (eg. bluetooth, ZigBee or Wifi modules) making it perfect for prototyping HCI devices. In the context of TUIs, the FreeIMU framework allows to enrich with orientation and motion sensing capabilities any object. Such objects could become virtual information handles which allow the user to move or rotate their virtual representation by interacting with them.

\section{FreeIMU hardware}

We designed the FreeIMU hardware following an iterative design process: in less than two years, thanks to low cost PCB prototyping services and DIY soldering capabilities, we designed, produced and tested, in collaboration with users from the community, seven different board designs. This design process allowed us to try different sensors available on the market and to adapt to new ones with improved features when they became available.

\figuremacro{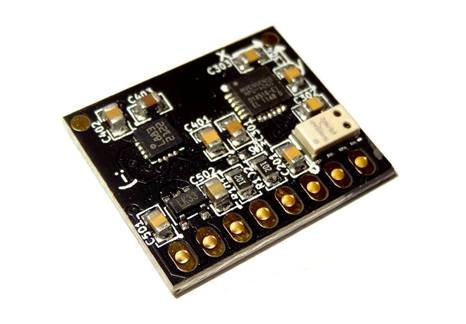}{A FreeIMU v0.4.3 board.}{}{0.5}

Our latest design, FreeIMU v0.4.3 (Figure \ref{freeimu_intro}), uses an MPU6050 gyroscope and accelerometer, an HMC5883L magnetometer and an MS5611-01BA high resolution atmospheric pressure sensor in a 22x19 mm sized board. The board is thus capable of measuring various input stimuli: 3-axis digital output angular rate sensor (gyroscope) with selectable full-scale range of $\pm250$, $\pm500$, $\pm1000$ and $\pm2000$ $\frac{\deg}{\sec}$, 3-axis digital output accelerometer with selectable full-scale range of $\pm2$, $\pm4$, $\pm8$ and $\pm16 g$, 3-axis digital output magnetometer with selectable full-scale range of $\pm0.88$, $\pm1.3$, $\pm1.9$, $\pm2.5$, $\pm4.0$, $4.7\pm$, $5.6\pm$ and $\pm8.1$ Gauss and an high resolution barometer with full scale range of 10 to 1200 mbar with selectable resolution of 0.065, 0.042, 0.027, 0.018, 0.012 mbar.
Such board can be bought ready to use for about 60 euro on some partner shops online.

The FreeIMU board can be connected to an Arduino or any other microcontroller through the I$^2$C bus with only four connections required. Additionally, sensors interrupt pins are available allowing advanced uses such as interrupt triggered readings or firmware actions.

All the hardware designs, including detailed bills of materials which are crucial when component selection has direct impact on the performance and reliability of the hardware, have been released under Open Hardware licences. This allows the designs to be studied, modified or embedded in other projects by third parties. Moreover, instead of using commercial hardware design software, generally expensive and rarely accessible to students, artists or hobbyists, we designed FreeIMU using KiCAD\footnote{\url{http://www.kicad-pcb.org/}}, an open source hardware design software.

\section{FreeIMU library}

The FreeIMU library contains all the software to operate a FreeIMU connected to an Arduino. The library, written in C/C++ with host software in Java/Processing and Python, contains high level functions to access raw or calibrated data from the sensors, perform sensor fusion algorithms as well as calibration procedures. 

Instead of spending months studying sensors functioning, sensor fusion algorithms or calibration problems, developers can simply obtain sensor data using high level APIs such as \verb|getRawValues()| which simply returns raw sensor data, \verb|getValues()| to access calibrated sensor readings as well as \verb|getQ()|, \verb|getEuler()| or \verb|getYawPitchRoll()| to obtain orientation estimates. Low level access to sensor configurations, whenever necessary by the developer, is provided by easy to use sensors C++ classes which avoid diving into the complex hardware aspects of the sensors. Developers are also capable of accessing data from the sensors through a simple serial communication protocol, useful when no custom firmware is needed but only the data from the sensors is required.\\
As an example of the FreeIMU library simplicity, listing \ref{inertial_mouse} implements an USB Inertial Mouse which uses pitch and roll angles to move the mouse pointer on screen.

\lstset{language=C,caption={Implementation of a simple USB Inertial Mouse based upon the FreeIMU framework and an Arduino Leonardo. Variables declaration omitted for brevity.},label=inertial_mouse}
\begin{lstlisting}
void loop() {

  my3IMU.getYawPitchRoll(ypr);
  
  // scale angles to mouse movements.
  int x = map(ypr[1], -90, 90, -10, 10);
  int y = map(ypr[2], -90, 90, -10, 10);
  
  // move mouse
  Mouse.move(-x, y, 0);
}
\end{lstlisting}

Regarding sensor fusion algorithms, in order to compute the orientation of the board, we make use of the algorithm by Mahony et al. \cite{mahony01}, originally developed for usage on Unmanned Aerial Vehicles (UAV), and later extended by Madgwick to incorporate the magnetic distortion compensation algorithms from his algorithm \cite{Madgwick2011} released as a reference quaternion implementation\footnote{\url{http://x-io.co.uk/node/8}}. Such algorithm, considerably lighter than traditional Kalman filters, allows us to obtain orientation data up to 400Hz even on the limited memory and computing power of the Arduino microcontroller.

The altitude of the board is computed by fusing the high resolution atmospheric pressure readings coming from the MS5611 barometer with gravity compensated accelerometer readings through a complementary filter. We are capable of measuring the device altitude with about 10cm precision and almost immediate response to motion. The planar displacements with regard to the Earth surface can be computed by double integrating the dynamic acceleration readings obtained from the accelerometer: such method yields imprecise results due to the noisy readings from the accelerometer and numerical errors during the integration, however it is still useful when the motion tracking does not need to be accurate.\\
When the sensor board is rigidly attached to the object whose orientation is being tracked, the library is also capable of detecting taps and double taps as well as their direction and magnitude.

Recently, the FreeIMU library has been extended with calibration routines which reduce or nullify errors in the orientation estimation induced by sensor biases. Through a guided procedure and by performing a set of predefined motions on the board, the user can obtain the calibration parameters which will then be used in the sensor software. A properly calibrated FreeIMU framework, is capable of sensing orientations inferior to 2 degrees without drifting.

The library is also accompanied by some visualization tools, written in Processing and Python, which allow the user to visualize the orientation and motion data from the sensors. These tools also serve as examples of interfacing with an host application on a PC. 

While the main platform to run a FreeIMU is an Arduino (8-bit ATMEGA AVR microcontroller), the community is constantly porting the library to other platforms: at the moment there are ports available for the PIC24 microcontroller and NXP LPC1343 Cortex M3 while work is being done to port it to other 32 bits platforms like the Arduino DUE or Teensy 3.0.

The FreeIMU library is released as open source software under the GPL v3 license and is available from the FreeIMU project website\footnote{\url{http://www.varesano.net/projects/hardware/FreeIMU}}.

\section{FreeIMU in Research, Education and Hobbies}
\label{freeimu_third}

Despite that the FreeIMU framework has been released only one and a half year ago, it has been already adopted in many research, education and hobbyists projects.

In \cite{6229434}, Calore et al. used a FreeIMU and its library to extend a standard camera with orientation sensing capabilities. This enabled them to develop algorithms to automatically use the orientation data to correct pictures for horizon and keystone effect distortions.

Varesano and Vernero used the FreeIMU framework as key prototyping tool during the development of PALLA\cite{Varesano:2012:IPN:2367616.2367621}, a spherical wireless input device designed for leisure activities for young children or elder users which uses the orientation and motion sensing capabilities provided by FreeIMU as the main input features of the device. Such input features have been used in prototyping Pandagolf, a TUI based golf video game.

The Pentacle \cite{Pentacle2012}, by Mitchell et al., is instead ``a wireless interface for new musical expression'': by using a FreeIMU connected to an Arduino FIO\footnote{\url{http://arduino.cc/en/Main/ArduinoBoardFio}} with vibration sensing microphones and wireless communication, the authors developed a musical instrument of dodecahedron shape. With the device, musicians can physically interact with digitally generated music through an orientation and motion sensing based interaction as well as through touching the microphone-equipped surfaces of the device.

In \cite{Patillo2012,Patillo2011}, Patillo developed the SmartSkelethon: starting from FreeIMU Open Hardware designs and library, the author extended the board and software to allow the daisy chaining of many IMUs so that they could be attached to the limbs of a teaching skeleton. The SmartSkelethon thus allows students of the Human Anatomy and Physiology classes to interact with a virtual skeleton in order to visualize the effects of limb motions into muscles and joins attempting to ease the understanding and memorization of them.

Russel et al instead, starting from the FreeIMU library and using other open source hardware, developed a low-cost IMU based upon Jeenode, an Arduino compatible board with wireless communication capabilities\cite{6229719}. This platform has been used as prototyping platform for developing stride time estimation algorithms for usage in the anticipatory prediction of falls for older users\cite{6226657}.

Veresano et al used the FreeIMU Framework as teaching tool in an Orientation and Motion sensing focused studio\cite{Varesano_tei2013_studio}. The same authors are also extending the FreeIMU Framework with wireless communication and a CPU on board to be used as prototyping tool of unconventional HCI devices making use of DIY and Personal Fabrication technologies\cite{Varesano_tei2013_doctoral}.

Reifberger and Br\"uller, students at the University of Applied Sciences - Upper Austria, are using FreeIMU to extend a 3D head mounted display with orientation and motion sensing capabilities to allow for a completely immersive virtual reality gaming experience\footnote{\url{http://magma-software.blogspot.com/}}.

The FreeIMU has been also used by many hobbists in various kinds of projects: virtual reality, games, robots, etc. FreeIMU has been extensively tested also as main sensor hardware in quadcopter UAVs applications: both the MultiWii\footnote{\url{http://www.multiwii.com/}} and Megapirate NG\footnote{\url{http://code.google.com/p/megapirateng/}}, hobby grade open source flying controller software, support the FreeIMU hardware on Arduino based platforms.

\section{Related Works}

While there is a wide literature that proposes frameworks to interact with commercial controllers like the WiiMote (eg: \cite{Schlomer:2008:GRW:1347390.1347395}) and general purpose sensor nodes (eg. \cite{Benbasat:2005:CMW:1147685.1147752}) there are only few works proposing frameworks composed by both hardware and software libraries for prototyping orientation and motion sensing. In \cite{Benbasat:2001:IMF:647592.728869} the authors propose a framework composed by a wireless sensor board and gesture recognition algorithms. While their work on the algorithms is still interesting, their sensor hardware is now considerably outdated and the sensors used in FreeIMU are considerably more accurate, faster and easier to interface. 

\textit{DUL Radio}\cite{Brynskov:2012:DTS:2148131.2148178} is a small sensor board containing an accelerometer, a microcontroller and wireless communication which has been designed for ease-of-use. While this device may be interesting for inexperienced developers, it may be also constraining compared to Arduino, as also noted by the authors. Our framework also aims to ease sensor based interaction prototyping, however we concentrated on not loosing the flexibility and extensibility of the Arduino platform while providing wider sensor input.

Concerning commercial projects similar to our, when we started working on FreeIMU back in 2010, there wasn't any similar board using all digital sensors. That's actually why the development of FreeIMU started. At the moment however, some prototype friendly boards containing 9 DOM MARG sensors are available. Sparkfun 9 DOF Sensor Stick \footnote{\url{https://www.sparkfun.com/products/10724}}, Razor IMU\footnote{\url{https://www.sparkfun.com/products/10736}} or 3D Robotics ArduIMU+ v3\footnote{\url{http://store.diydrones.com/ArduIMU_V3_p/kt-arduimu-30.htm}} are example of similar commercial products which also are Open Hardware and some are equipped with minor software libraries. 

Vectornav VN-100 \footnote{\url{http://www.vectornav.com/products/vn100-rug}} and X-IO X-IMU \footnote{\url{http://www.x-io.co.uk/node/9}} are instead remarkable examples of closed source commerial IMUs since they provide ready to use proprietary orientation sensing algorithms.

From an hardware point of view, the FreeIMU hardware is superior to both the Sparkfun 9 DOF Sensor Stick and Razor IMU since the sensors used in the FreeIMU v0.4.3 board are more precise and less noisy\footnote{The ADXL345 accelerometer used in the Sparkfun 9 DOF Stick and Razor IMU only gives 256 digits/g of resolution ($\pm2g$ scale) while the MPU6050 of the FreeIMU v0.4.3 gives 16,384 digits/g ($\pm2g$ scale) thus being much more precise. Same considerations are applicable to the gyroscope.}. The FreeIMU also features a barometer allowing to track  the altitude of the prototyped device. The Razor IMU is also limited by being clocked at 8MHz while the FreeIMU is compatible with 16MHz Arduino boards. The ArduIMU+ v3 uses sensors similar to those of the FreeIMU but doesn't provide a barometer. Arguably the Razor IMU and ArduIMU+ v3 may be considered easier to use since they provide sensors and CPU on the same board while the FreeIMU needs to be connected to an host Arduino board: we believe however that this is not a limitation when prototyping HCI devices since in the prototype there will probably be other sensors or input devices requiring additional connections, thus the 4 connections required by the FreeIMU board shouldn't be a problem.

From a software point of view, the FreeIMU library provides a much richer and polished API compared to those offered by 3rd parties on the Razor IMU or ArduIMU+ v3. The FreeIMU library is also the only one offering altitude estimation and a calibration GUI. While we don't currently have tests comparing the FreeIMU library orientation algorithm performance with the software for the Razor IMU or ArduIMU+ v3, according to some users ``\textit{the FreeIMU implementation smoother, less noisy, and more responsive than the DCM based ArduIMU code}''\footnote{http://diydrones.com/forum/topics/freeimu-firmware-on-arduimu}: this is the reason why some many Razor IMU or ArduIMU+ v3 users ended up using or porting the FreeIMU library to their hardware instead of using their board software.

Compared to closed source IMU platforms, such as the Vectornav VN-100 and X-IO X-IMU, while we don't currently have test comparing the orientation sensing of our platform with those, we believe that the FreeIMU project can still give much greater possibilities then using a commercial platform. By being completely open, both from an hardware and software point of view, it can be freely implemented in 3rd parties derivative projects or products, something not possible when using closed source IMU platforms.

The FreeIMU is also the only IMU project we know supported by a very active community of users, organized in the FreeIMU community website\footnote{http://freeimu.varesano.net/}, sharing projects, code improvements and helping each other.

\section{Conclusions and Future work}

This paper presented FreeIMU, a framework for prototyping orientation and motion sensing devices composed by a sensor board and a software library. We briefly explained the rationale behind the creation of the framework, we described its hardware and software components and shown various projects already taking advantage of the framework.

As explained, we believe that FreeIMU can represent a valuable contribution in HCI research since it allows for the easy prototyping of orientation and motion sensing capable devices abstracting from the complex aspects of hardware, algorithms and low level programming. The complete openness of the framework, both from the hardware and software points of view, supports the creation of derivative projects which can take advantage from the simplicity of the library. The flexibility of the Arduino platform, which runs the FreeIMU library software, also allows to further extend the capabilities offered by our framework with additional input, output or feedback.

We believe that the number of high quality derivative projects based upon the FreeIMU framework in the academic, education and hobbyists communities, as presented in section \ref{freeimu_third}, is a proof of the quality and possibilities offered by our framework.

As future work, from a software point of view we are currently interested in adding gesture recognition to our library as well as improving the calibration procedure to ease the process for the end user of the prototype. The research in inertial sensors calibration is quite active so its probable that better calibration algorithms will be published in the future. From an hardware point of view, we are continuing to follow the development in sensors as we want to keep our board update with newer sensors in the future. We are also investigating the possibility of further extending the hardware adding a microcontroller and a wireless communication module, creating an Arduino-like with on board sensors and communication to further ease the prototyping.

\bibliographystyle{splncs03}
\bibliography{references}

\end{document}